\documentclass[twocolumn,times]{aastex61}

\journalinfo{Published as Ebeling et al., 2018, ApJ, 852, L7}

\shorttitle{eMACS1341-QG-1}
\shortauthors{Ebeling et al.}

\begin{document}

\title{Thirty-fold: Extreme gravitational lensing of a quiescent galaxy at $z=1.6$}

\correspondingauthor{Harald Ebeling}
\email{ebeling@ifa.hawaii.edu}

\author{H. Ebeling}
\affil{Institute for Astronomy, University of Hawai`i, Honolulu, HI 96822, USA}

\author{M. Stockmann}
\affil{Cosmic Dawn Center, Niels Bohr Institute, University of Copenhagen, Juliane Maries Vej 30, K{\o}benhavn {\o}, 2100, Denmark}

\author{J. Richard}
\affil{Univ Lyon, Univ Lyon 1, Ens de Lyon, CNRS, Centre de Recherche Astrophysique de Lyon UMR5574, F-69230 Saint-Genis-Laval, France}

\author{J. Zabl}
\affil{IRAP, Institut de Recherche en Astrophysique et Plan\'{e}tologie, CNRS, France/Universit\'{e} de Toulouse, UPS-OMP, 9, avenue du Colonel Roche, F-31400 Toulouse, France}

\author{G. Brammer}
\affil{Space Telescope Science Institute, 3700 San Martin Drive, Baltimore, MD 21218, USA}

\author{S. Toft}
\affil{Cosmic Dawn Center, Niels Bohr Institute, University of Copenhagen, Juliane Maries Vej 30, K{\o}benhavn {\o}, DK-2100, Denmark}

\author{A. Man}
\affil{European Southern Observatory, Karl-Schwarzschild-Strasse 2, D-85748 Garching bei M\"unchen, Germany}

\begin{abstract}
We report the discovery of eMACSJ1341-QG-1, a quiescent galaxy at $z=1.594$ located behind the massive galaxy cluster eMACSJ1341.9$-$2442 ($z=0.835$). The system was identified as a gravitationally lensed triple image in Hubble Space Telescope images obtained as part of a snapshot survey of the most X-ray luminous galaxy clusters at $z>0.5$ and spectroscopically confirmed in ground-based follow-up observations with the ESO/X-Shooter spectrograph. From the constraints provided by the triple image, we derive a first, crude model of the mass distribution of the cluster lens, which predicts a gravitational amplification of a factor of $\sim$30 for the primary image and a factor of $\sim$6 for the remaining two images of the source, making eMACSJ1341-QG-1 by far the most strongly amplified quiescent galaxy discovered to date. Our discovery underlines the power of SNAPshot observations of massive, X-ray selected galaxy clusters for lensing-assisted studies of faint background populations.
\end{abstract}


\section{Introduction} \label{sec:intro}

Understanding the origin and evolutionary history behind the diverse galaxy population encountered in the local Universe is one of the most pressing goals of present-day astronomy. In the course of the past decade, two highly complementary populations have received particular attention: the first generations of galaxies to have formed at redshifts $z{=}$6--10 \citep[e.g.,][]{ Bunker2010,Yan2011,Bouwens2012,Bouwens2015,Oesch2012}, and massive quiescent galaxies at $z{=}$1.5--4 as  precursors of the local population of early-type galaxies \citep[e.g.,][]{Franx2003,vanDokkum2006, Newman2012,Glazebrook2017}. However, in both regimes, the faintness of all but the most luminous representatives of these high-redshift populations has created major observational challenges. 

Gravitational lensing has proven invaluable in this context, as it can boost the brightness of suitably positioned background galaxies by up to, and in rare cases over, an order of magnitude \citep[e.g.,][]{Yuan2012,Zheng2012,Coe2013, Atek2015,Finkelstein2015,McLeod2015}, thereby allowing us to study average high-redshift galaxies that would be firmly beyond our observational reach without lensing assistance. Since lensing indiscriminately affects all fortuitously aligned background objects, the vast majority of lensed galaxies, at any redshift, are, however, late-type systems.

As an immediate consequence, no observational study of a quiescent galaxy at $z{=}$1.5--4 has, so far, benefited from the extreme magnifications (by factors of 30 or more) that have allowed detailed investigations into the properties of distant star-forming galaxies \citep[e.g.,][]{Smail2007,Swinbank2010}. Even at much more modest magnifications ($\mu$) of a factor of a few, lensing-assisted studies of massive early-type galaxies at $z{\sim}2$ remain rare, recent examples being RG1M0150 at $z=2.64$ \citep[$\mu_{\rm max}=3.9$,][]{Newman2015} and MACS2129-1 at $z=2.15$ \citep[$\mu_{\rm max}=4.6$,][]{Toft2017}, both gravitationally boosted by MACS clusters \citep{Ebeling2001,Ebeling2007,repp2017}. 
 In spite of the modest lensing amplification, these observations enabled studies at high resolution (both spatial and spectral) of representatives of a galaxy population believed to play a crucial role in the evolutionary path of passively evolving systems \citep[e.g.,][]{Toft2012,vandeSande2013,Newman2015}. While it is widely accepted that already at $z{\sim}1.5$ a majority of the most massive galaxies had evolved stellar populations and form few stars, the observational evidence behind this picture is not conclusive, in particular regarding the puzzlingly compact size of some of these galaxies, the quenching mechanism, and the impact of dust on the apparent prominence of the old stellar population. 

In this Letter, we describe the discovery of a galaxy that holds great promise for fresh observational insights into the origin and evolution of early-type galaxies: an extremely magnified quiescent galaxy at $z=1.59$ lensed by the massive galaxy cluster eMACSJ1341.9--2442. A companion paper (A.\ Man et al., in preparation) discusses the galaxy's properties.\\

\section{eMACSJ1341.9--2442} \label{sec:e1341}

Discovered in the course of the extended Massive Cluster Survey \citep[eMACS;][]{Ebeling2013}, which aims to identify extremely massive galaxy clusters at $z{>}0.5$ among the faintest X-ray sources detected in the ROSAT All-Sky Survey \citep{Voges1999}, eMACSJ1341.9--2442 was established in 2013 as a very X-ray luminous galaxy cluster at $z{=}0.835$ ($L_{\rm X}{=}1.6\times10^{45}$ erg s$^{-1}$, 0.1--2.4 keV). Unambiguous confirmation of a very massive system was obtained via ground-based follow-up observations with Gemini-N/GMOS and Keck-2/DEIMOS that showed the cluster to be optically rich, far from relaxed, and featuring an exceptionally high cluster velocity dispersion of 1700$_{-290}^{+190}$ km s$^{-1}$ (based on 23 redshifts) with no evidence of significant substructure along our line of sight. The extreme velocity dispersion of eMACSJ1341.9--2442 is either indicative of an exceptionally high mass or of bulk motions along the line of sight induced by an ongoing merger event, as observed in, e.g., A370 \citep{Lagattuta2017} or MACSJ0553.4--3342 \citep{Ebeling2017}. A more extensive discussion of eMACSJ1341.9--2442 is provided by H.\ Ebeling et al.\ (2017, in preparation) as part of an overview of the eMACS cluster sample.  

\section{An exceptional strong-lensing event} \label{sec:hst}

\begin{figure}
\includegraphics[width=0.47\textwidth,clip,trim=41mm 26mm 38mm 29mm]{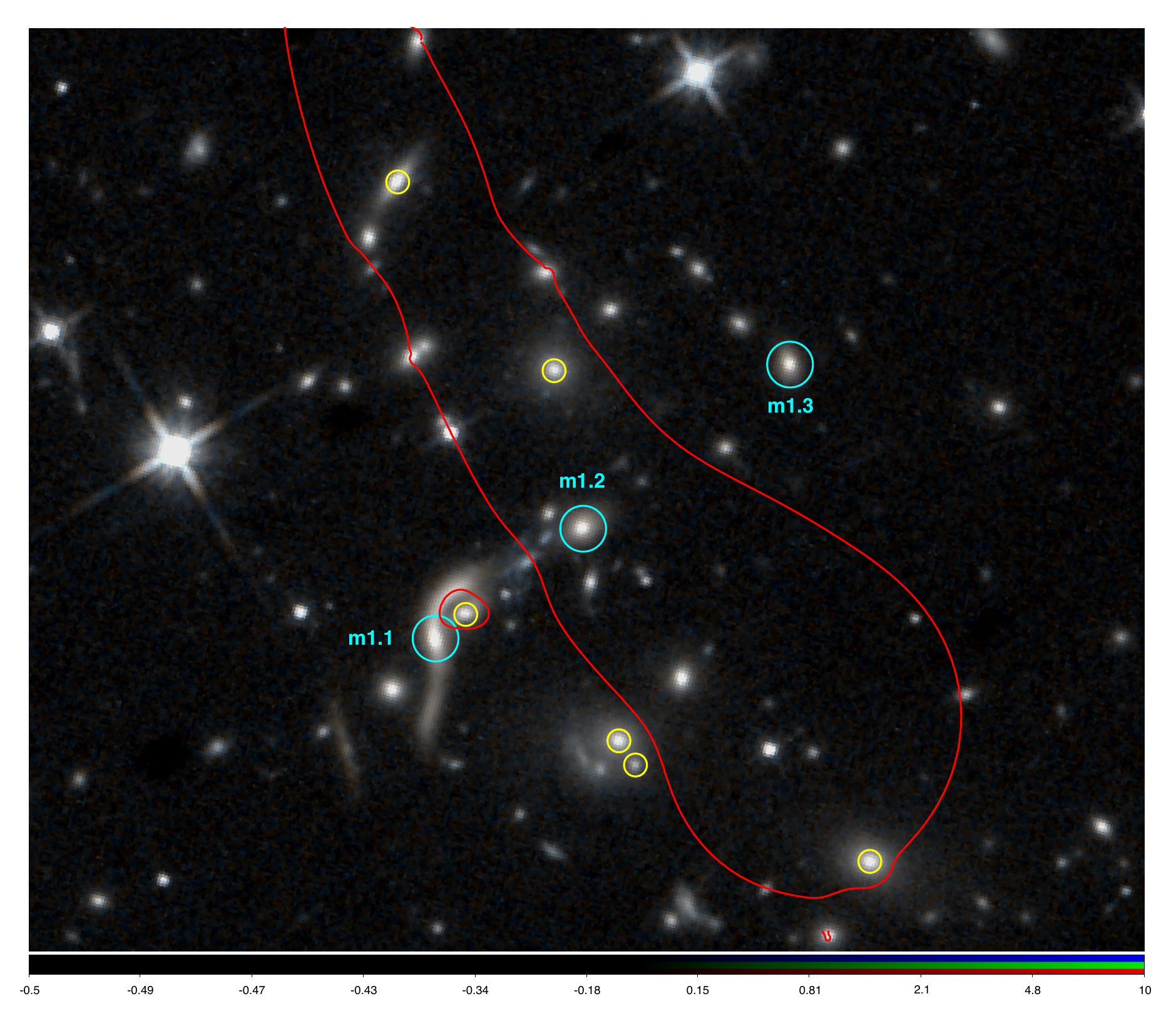}
\caption{WFC3 image (F110W, F140W; shown area: $30{\times}30$ arcsec$^2$) of the core of eMACSJ1341.9--2442, obtained for GO-14098 on 2016 September 2.  Spectroscopically confirmed cluster members are marked in yellow. The three primary components of the gravitationally lensed multiple-image system are marked and labeled in cyan; the strongly curved arc in image m1.1 is largely due to galaxy--galaxy lensing by a cluster member. The critical line for lensing of an object at $z{=}1.6$ is shown in red.\label{fig:hst}}
\end{figure}

\subsection{Observations}

eMACSJ1341.9--2442 was observed with the Wide-Field Camera 3 (WFC3) on board the Hubble Space Telescope (HST) in the F110W and F140W passbands (706 s each) in the course of our Cycle 23 SNAPshot program GO-14098 (PI: Ebeling). The WFC3 image shown in Fig.~\ref{fig:hst} reveals a dramatic example of strong gravitational lensing in the form of a bright triple-image system, the brightest member of which is also lensed by an individual cluster galaxy, creating a complex s-shaped arc (m1.1 in Fig.~\ref{fig:hst}). 

\begin{figure}
\includegraphics[width=0.47\textwidth]{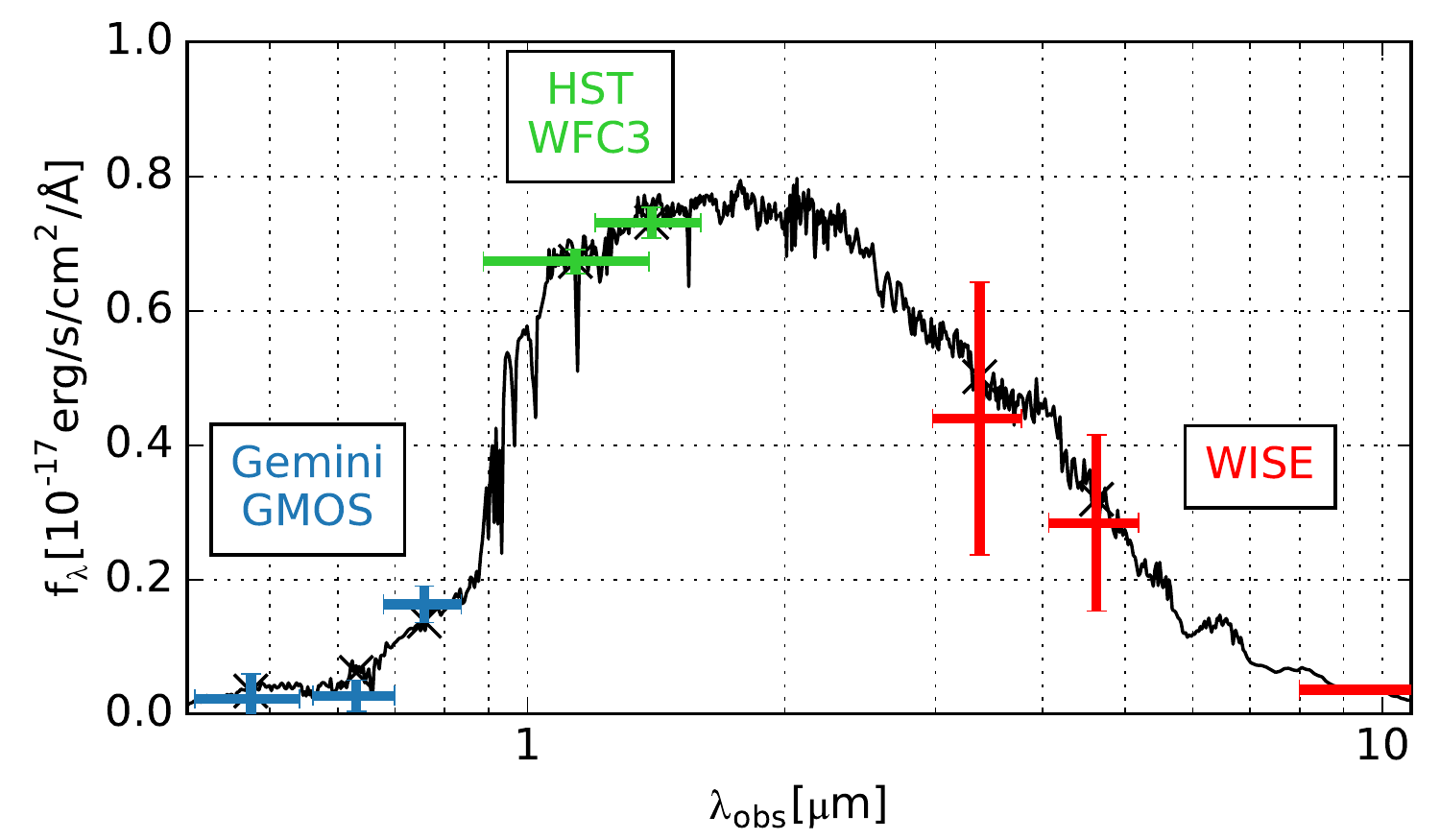}
\caption{Best-fit SED model ($z{=}$1.35) based on aperture photometry for image m1.1; except for a lower normalization, the SEDs of m1.2 and m1.3 are fully consistent. Blending in the \textsl{WISE} data is accounted for by setting the respective uncertainties to 0.5 mag.  \label{fig:sed}}
\end{figure}

Prompted by this discovery, we attempted to model the spectral energy distribution (SED) for image m1.1,\footnote{Although the cyan circle in Fig.~\ref{fig:sed} encloses only the brightest part of image m1.1, the S-shaped arc extending to either side of it is part of this image, representing different parts of the lensed galaxy in the source plane, and is included in the photometric aperture.} by combining photometry from HST/WFC3 (F110W and F140W), Gemini/GMOS (\textsl{g}$^\prime$, \textsl{r}$^\prime$, and \textsl{i}$^\prime$; program GN-2015A-Q-25, PI: Ebeling), and the \textsl{Wide-field Infrared Survey Explorer} \citep[\textsl{WISE};][]{Wright2010,Mainzer2011}. All data were found to be well described by the SED of a quiescent galaxy at $z\sim 1.35$ (Fig.~\ref{fig:sed}).

\begin{figure}[t]
\hspace*{0mm}{\includegraphics[width=0.47\textwidth]{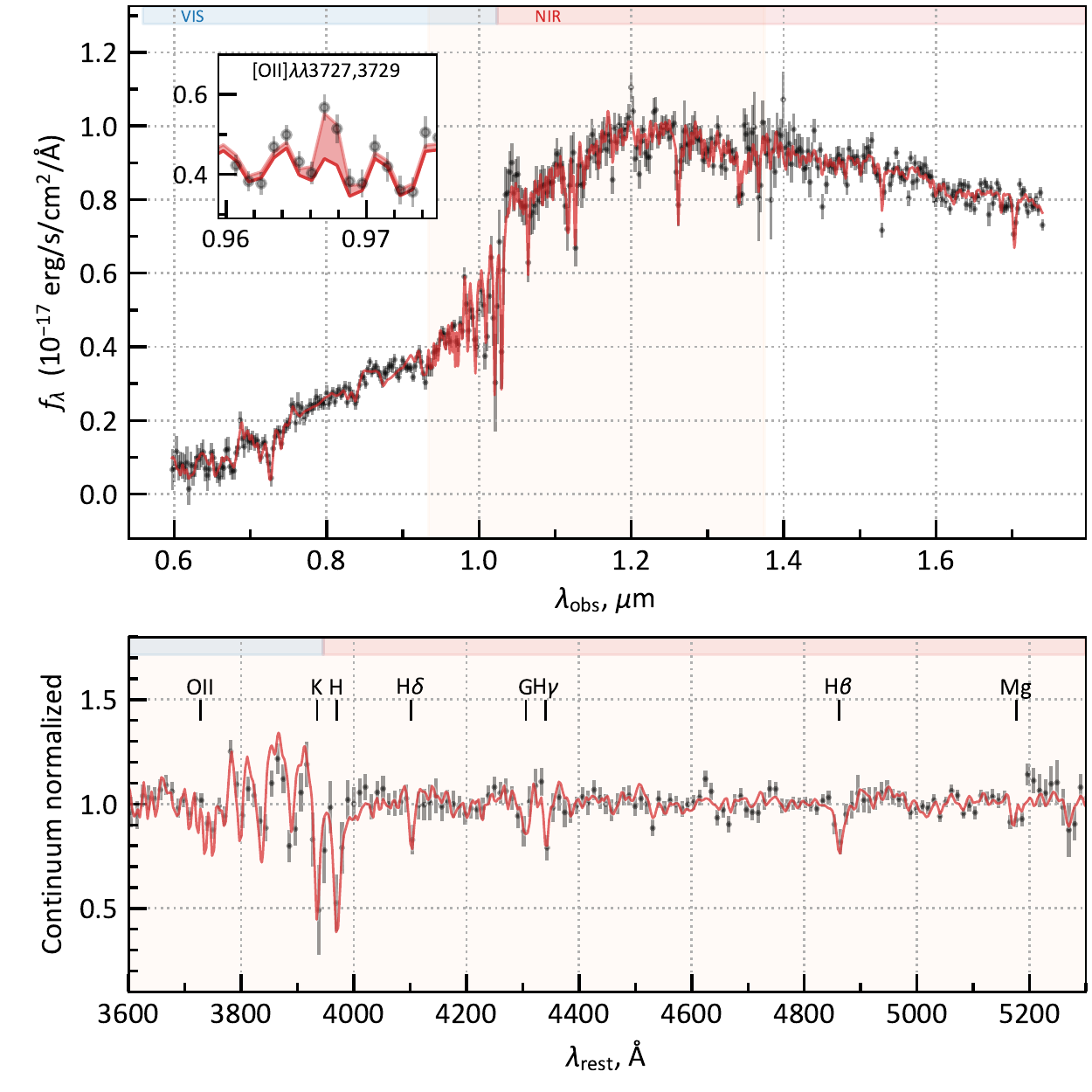}}
{\caption{Spectrum of image m1.1 from Fig.~\ref{fig:hst} as obtained with X-Shooter in 2017 July/August for program 099.B-0912(A) (PI: Stockmann), confirming the lensed object as a quiescent galaxy at $z=1.594$ (the inset in the upper panel shows the detection of a weak [\ion{O}{2}]$\lambda\lambda$3727,3729 emission line). Individual absorption features, such as the pronounced Ca K+H lines and the Balmer series in absorption, are labeled in the lower panel, which shows the spectrum in the galaxy rest frame after normalization by the continuum.\label{fig:xshooter}}}
\end{figure}

Based on the evidence presented so far, we were awarded a 4-hr observation of m1.1 with the X-Shooter spectrograph on the VLT (program 099.B-0912(A), PI: Stockmann), three hours' worth of which were performed in 2017 July/August. The high-quality spectrum from X-Shooter, shown in Fig.~\ref{fig:xshooter}, establishes the redshift of eMACSJ1341-QG-1 as $z=1.594$ and identifies the system unambiguously as quiescent and dominated by an old stellar population. Preliminary spectral fits with the Prospector fitting code\footnote{publicly available at:  \url{https://github.com/bd-j/prospector}} \citep[][B.\ Johnson et al.\ 2017, in preparation]{Leja2017} and FSPS models \citep[flexible stellar population synthesis;][]{Conroy2009,Conroy2010} yield an SSP-equivalent age of 1.5 Gyr, modest dust reddening ($A_V \approx 0.2$), and a short timescale of star formation ($\tau < 0.1$ Gyr). Along with the low [\ion{O}{2}] equivalent width, these constraints on the star-formation history suggest a very low specific star-formation rate ($\mathrm{sSFR}<0.01$ Gyr$^{-1}$), placing the star formation in eMACSJ1341.9$-$2442 one to two orders of magnitude below that of typical star-forming galaxies at $z\sim 1.5$ \citep[e.g.,][]{Whitaker2012}. An in-depth analysis of the X-Shooter spectrum is presented in Man et al.\ (2017, in preparation).

\subsection{Lens Model}

Using \textsc{LensTool} \citep{Jullo2007}, we derive a first, crude lens model from the constraints provided by the three images of eMACSJ1341-QG-1, adopting for all of them the spectroscopic redshift of $z=1.594$ measured for image m1.1, and including all spectroscopically confirmed cluster members as small-scale perturbers \citep[for a detailed description of our modeling approach see][and references therein]{Richard2014}.

The complexity of the mass distribution of the cluster lens (as indicated by the extended galaxy distribution and the absence of a single cluster core), combined with the limited number of constraints provided by our sole multiple-image system, creates a challenge for the creation of a credible first lens model. A model based on two mass concentrations with additional galaxy-scale perturbers does, however, reproduce the locations of the three images of system m1 to better than 1\arcsec\ accuracy. We show the corresponding critical lines for strong lensing of a source at $z=1.6$ in Fig.~\ref{fig:hst}. Although our description of the mass distribution in the cluster lens and hence of the magnification map is highly uncertain for much of the cluster, all plausible models explored by us agree reasonably well on the gravitational magnification at the locations of images m1.1, m1.2, and m1.3. We predict modest magnifications of $\mu=6\pm 2$ for images m1.2 and m1.3, which, however, still exceed those experienced by RG1M0150 \citep{Newman2015} and MACS2129-1 \citep{Toft2017}. For image m1.1 our model predicts a dramatic magnification factor of $\mu=30\pm 8$ (both for its brightest part and its full extent including the S-shaped arc), much higher than the magnifications of any previously known quiescent galaxy in this redshift regime. These values are in agreement with the brightness of image m1.1 and the flux ratios between all three images and
imply a lensing-corrected, total absolute magnitude of M$_{\rm V}{\sim}-21.2$ for our target galaxy, over 1 mag fainter than $M^\ast$ at $z{=}$1.1--1.5 \citep{Marchesini2012}. 

The identification of additional strong-lensing features will be critical to improve our current lens models and thus allow (a) a robust determination of the mass distribution across the core of eMACSJ1341.9--2442 and (b) a faithful reconstruction of the morphology of eMACSJ1341-QG-1 in the source plane.

\section{Summary}

Owing to the presence of the foreground cluster eMACSJ1341.9--2442 ($z=0.835$), the  quiescent galaxy eMACSJ1341-QG-1 at $z=1.594$ can be studied at an unprecedented magnification of $\sim$30, providing insights into the properties and evolution of the precursors of early-type galaxies in the local universe. Although ground-based spectroscopy with facilities in Chile and on Maunakea will allow the characterization of the stellar populations of eMACSJ1341-QG-1, an analysis of its spatial profile and any radial dependencies of its properties relies on the availability of resolved colors and a robust lens model that allows the reconstruction of the galaxy in the source plane. HST imaging in multiple filters will be critically important to achieve either of these goals.

\acknowledgments
H.E.\ gratefully acknowledges financial support from NASA ADAP grant NNX11AB04G as well as funding from STScI for HST program GO-14098; S.T.\ and M.S.\ acknowledge support from the ERC Consolidator Grant funding scheme (project ConTExt, grant number 648179). The Cosmic Dawn Center is funded by the Danish National Research Foundation. J.R.\ acknowledges support from ERC starting grant 336736-CALENDS. J.Z.\ acknowledges support of the OCEVU Labex (ANR-11-LABX-0060) and the A*MIDEX project (ANR-11-IDEX-0001-02) funded by the ``Investissements d'Avenir" French government program managed by the ANR.

Based on observations made with the NASA/ ESA Hubble Space Telescope, obtained at the Space Telescope Science Institute, which is operated by the Association of Universities for Research in Astronomy, Inc., under NASA contract NAS 5-26555. 

Based on observations collected at the European Organisation for Astronomical Research in the Southern Hemisphere under ESO programme 099.B-0912(A).

Data presented herein were obtained at the W.M.\ Keck Observatory, which is operated as a scientific partnership among the California Institute of Technology, the University of California, and the National Aeronautics and Space Administration. The observatory was made possible by the generous financial support of the W.M.\ Keck Foundation. 

Based on observations obtained at the Gemini Observatory, which is operated by the Association of Universities for Research in Astronomy, Inc., under a cooperative agreement with the NSF on behalf of the Gemini partnership: the National Science Foundation (United States), the National Research Council (Canada), CONICYT (Chile), Ministerio de Ciencia, Tecnolog\'{i}a e Innovaci\'{o}n Productiva (Argentina), and Minist\'{e}rio da Ci\^{e}ncia, Tecnologia e Inova\c{c}\~{a}o (Brazil).

This publication makes use of data products from the \textsl{Wide-field Infrared Survey Explorer}, which is a joint project of the University of California, Los Angeles, and the Jet Propulsion Laboratory/California Institute of Technology, and NEOWISE, which is a project of the Jet Propulsion Laboratory/California Institute of Technology. \textsl{WISE} and NEOWISE are funded by the National Aeronautics and Space Administration.

The authors wish to recognize and acknowledge the very significant cultural role and reverence that the summit of Maunakea has always had within the indigenous Hawaiian community.  We are most fortunate to have the opportunity to conduct observations from this mountain.

\vspace{5mm}
\facilities{HST (WFC3), Keck (DEIMOS), Gemini-N (GMOS, program GN-2015A-Q-25we), VLT (X-Shooter, program 099.B-0912(A))}

\end{document}